\documentclass[pra,twocolumn]{revtex4-1} 
\usepackage{braket,amsmath,amssymb}
\usepackage{mathtools}
\usepackage{mathbbol}
\usepackage{tikz}
\usepackage{verbatim}
\usetikzlibrary{matrix,decorations.pathreplacing, calc, positioning}
\usepackage{color}
\usepackage{enumitem}
\usepackage{graphicx}
\usepackage{dcolumn}
\usepackage{bm}
\usepackage{amsfonts}
\usepackage{float}
\usepackage{mathrsfs}
\usepackage{color,hyperref}
\definecolor{darkblue}{rgb}{0.0,0.0,0.4}
\hypersetup{colorlinks,breaklinks,
linkcolor=darkblue,urlcolor=darkblue,
anchorcolor=darkblue,citecolor=
darkblue,pdfauthor=Chandan, pdftitle=ChandanCVMeasurement }

\newcommand{\ie}{\textit{i}.\textit{e}.}

\begin{document}
\title{Estimation of Wigner distribution of single
mode Gaussian states: a comparative study}
\author{Chandan Kumar}
\email{chandan.quantum@gmail.com}
\affiliation{Department of Physical Sciences,
Indian
Institute of Science Education and
Research Mohali, Sector 81 SAS Nagar,
Punjab 140306 India.}
\author{Arvind}
\email{arvind@iisermohali.ac.in}
\affiliation{Department of Physical Sciences,
Indian
Institute of Science Education  and
Research Mohali, Sector 81 SAS Nagar,
Punjab 140306 India.}
\begin{abstract}
In this work, we consider the estimation of
single mode Gaussian states using four different measurement
schemes namely: i) homodyne measurement, ii)
sequential measurement, iii) Arthurs-Kelly scheme, and iv)
heterodyne measurement, with a view
to compare their relative performance. To this end, we work in
the phase space formalism, specifically at the covariance matrix
level, which provides an elegant and intuitive way to
explicitly carry out involved calculations.  We show that
the optimal performance of the Arthurs-Kelly scheme and the
sequential measurement is equal to the heterodyne
measurement.  While the heterodyne measurement outperforms
the homodyne measurement in the mean estimation of squeezed
state ensembles, the homodyne measurement outperforms the
heterodyne measurement for variance estimation of squeezed
state ensembles up to a certain range of the squeezing parameter.
We then modify the Hamiltonian in the
Arthurs-Kelly scheme, such that the two meters can have
correlations and show that  the optimal performance
is achieved when the meters are
uncorrelated. We expect that these results will be useful
in various quantum information and quantum
communication protocols.
\end{abstract}
\maketitle
\section{Introduction}
Reconstruction of quantum states by performing measurements
on an ensemble of systems prepared in identical but
unknown states is known as quantum
state estimation (QSE) or quantum state tomography
(QST)~\cite{james-pra-2001,paris-2011,lvovsky-rmp}.  QSE is
 important in quantum mechanics
and quantum information processing, and finding schemes for
its efficient execution is an active area of
research~\cite{vallone-prl-2016,vallone-prl-2018,walmsley-pra-2020}.
Ideal QSE requires infinite copies of a quantum system,
which is impractical in the real world. Usually, an
experimentalist is provided with a fixed number of
identically prepared systems, and thus one would like to
know the advantages and limitations of various measurement
schemes and select the best scheme as per the requirements.
We consider several different measurement schemes that can
be employed for the QSE of continuous variable (CV) systems.

Homodyne measurement is one of the most widely employed
measurement scheme in CV systems, which measures either the
$\hat{q}$ quadrature or the $\hat{p}$ quadrature or any
other phase rotated quadrature
operator~\cite{yuen-83,abbas-83,schumaker-1984,banaszek-pra-1997}.
It has been shown that various quasiprobability
distributions such as Glauber-Sudarshan distribution, Wigner
distribution, and Husimi distribution~\cite{vogel-pra-1989}
can be estimated using measurements of the rotated
quadrature operators and has also been demonstrated
experimentally~\cite{raymer-prl-1993}.  One can also think
of sequential measurement (SM) of a pair of conjugate observables
in CV systems, where the observables are measured one after
another~\cite{jukka-pra-2009,lorenzo-prl-2013,da-jpa-2017}.
The sequential measurement of two non-commuting observables
has been employed to reconstruct the Moyal $M$ function,
which is the Fourier transform of the Wigner
function~\cite{lorenzo-prl-2013}.  
Arthurs and Kelly, in a unique effort
extended the von Neumann measurement
scheme to the joint measurement of two non-commuting
observables~\cite{arthurkelly-1964}. In this joint measurement
the consequence of non-commutativity is the introduction of additional
minimum noise in the outcomes of both the observables.
Similarly, a heterodyne measurement, which is equivalent to
an eight-port measurement (double-homodyne),  
can be employed for
the joint measurement of two non-commuting
observables~\cite{javan-62,read-1965,carleton-1968,gerhardt-1972,yuen-1980,
yuen-1982,arthurs-1982,shapiro-1984, shapiro-1985,
gardiner-1987, martens-pla-1991, raymer-apj-1994}. 
For the 
Hetrodyne measurement, the vacuum noise is added in both the
non-commuting observables and unlike in the Arthurs-Kelly scheme this
cannot be distributed among the observables at will. It
has been shown that the heterodyne measurement can do a
better estimation of the first and the second order moments
of the quadrature operators of Gaussian and non-Gaussian
states compared to the homodyne
measurement~\cite{rehack-scirep-2015,rehack-prl-2016,rehack-pra-2017}.
The superiority of localized phase space sampling with
unbalanced homodyne measurement~\cite{vogel-pra-1996}
compared to delocalized heterodyne measurement has also been
shown~\cite{teo-pra-2017}.

In this article, we provide a comparative study 
of the state estimation efficiency of
different measurement schemes including homodyne
measurement, sequential measurement, Arthurs-Kelly
measurement, and heterodyne measurement.
To this end, we
consider an ensemble of $N$ identically prepared single mode
Gaussian states.
Gaussian states are defined as states with a Gaussian Wigner
distribution function and we need to estimate the mean and the covariance
matrix to  completely
reconstruct such states. 
We  further assume that the Gaussian states are either
squeezed in $\hat{q}$ or the
$\hat{p}$ quadrature. Such states have been
 employed in squeezed state CV quantum 
key distribution
protocols~\cite{hillery-pra-2000,cerf-pra-2001,
grosshans_cvqkd} and are easier to estimate.

We provide analytical  expressions of the estimation
efficiency of the mean and the variance of an ensemble of
$N$ identically prepared Gaussian states.  We show that the
optimal performance of the Arthurs-Kelly scheme  and the
sequential measurement is equal to the heterodyne
measurement. 
For mean estimation, the heterodyne measurement
outperforms the homodyne measurement for squeezed coherent
state ensemble, but for variance estimation, the homodyne
measurement outperforms the heterodyne measurement for a
certain  squeezing parameter range. Then we proceed to a
modified Hamiltonian~\cite{busch-1985,bullock-prl-2013} in
the Arthurs-Kelly scheme that can entangle the two meters.
Here the results show that the optimal performance of the
scheme can only be obtained when the meters are
uncorrelated. Since the Hamiltonians involved in the
sequential measurement and the Arthurs-Kelly scheme  are
quadratic expressions in quadrature operators, the
corresponding symplectic transformations acting on the
quadrature operators or the phase space variables belong to
the real symplectic group $Sp(4,\mathcal{R})$ and
$Sp(6,\mathcal{R})$~\cite{arvind-pla-2020}, respectively. We
exploit this fact and explicitly work in phase space for
calculational simplicity.  We expect that these techniques
along with the results obtained in this work will be useful
in undertaking various studies in different quantum
information and quantum communication
protocols~\cite{Braunstein,weedbrook-rmp-2012}.

This article is organized as follows. In
Sec.~\ref{cvsystem}, we present the basic formalism of CV
system, while in Sec.~\ref{schemes}, we introduce various
measurement schemes and obtain the variance of probability
distributions corresponding to different quadrature
measurements.  In Sec.~\ref{results}, we analyze the
performance of various measurement schemes in  the mean and
the variance estimation.  In Sec.~\ref{modifiedak}, we
consider Arthurs-Kelly scheme with a modified interaction
Hamiltonian. Finally, in Sec.~\ref{conclusion}, we provide
some concluding remarks and discuss future prospects.
\section{Background}
\label{background}
In this section, we introduce the formalism of
$n$-mode CV system and describe various
measurement schemes.  
\subsection{Description of an $n$-mode CV system}
\label{cvsystem}
An $n$-mode continuous variable quantum system
can arise in different contexts, ranging from 
$n$-physical harmonic oscillators to $n$-modes 
of the electromagnetic field or of the lattice vibrations.
Although the description here is general, the physical system
that we have in mind is $n$-modes of the electromagnetic field.

An $n$-mode quantum CV system is described by $n$ pairs of Hermitian
quadrature operators, which can be collected in a vector form
as follows~\cite{arvind1995,Braunstein,weedbrook-rmp-2012,adesso-2014}:
\begin{equation}\label{eq:columreal}
\bm{\hat{ \xi}} =(\hat{ \xi}_i)= (\hat{q_{1}},\,
\hat{p_{1}} \dots, \hat{q_{n}}, 
\, \hat{p_{n}})^{T}, \quad i = 1,2, \dots ,2n.
\end{equation}
The canonical commutation relations can be
written compactly as (with $\hbar$=1):
\begin{equation}\label{eq:ccr}
[\hat{\xi}_i, \hat{\xi}_j] = i \Omega_{ij}, \quad (i,j=1,2,\dots,2n),
\end{equation}
where $\Omega$ is the symplectic form given by
\begin{equation}
\Omega = \bigoplus_{k=1}^{n}\omega =  \begin{pmatrix}
\omega & & \\
& \ddots& \\
& & \omega
\end{pmatrix}, \quad \omega = \begin{pmatrix}
0& 1\\
-1&0 
\end{pmatrix}.
\end{equation}

The Hilbert space of the $n$-mode system is spanned by the product
	 basis vector $ \vert n_1\dots n_i \dots n_n\rangle$ with $\{n_1,\, 
	\dots\,, n_i,\, \dots\,, n_n=0,\, 1, \dots ,\infty \} $, where the number $n_i$
	corresponds to photon number in the $i^{\text{th}}$ mode. The quantum states of this
	system are represented via density operators in this Hilbert
	space. The density operator $\hat{\rho}$ 
	is a Hermitian, non-negative and trace-one class operator.

We can alternatively describe an $n$-mode system in a
$2n$-dimensional phase space.
The Wigner distribution for a quantum system with density
operator $\hat{\rho}$ can be written as~\cite{wigner-1932}
\begin{equation}\label{eq:wig}
W(\xi) = (2 \pi)^{-n}\int d^n q'\, \langle
\underbar{q}-\frac{1}{2}
\underbar{q}^{\prime}|\hat{\rho}|\underbar{q}+\frac{1}{2}\underbar{q}^{\prime}
\rangle \exp(i \underbar{q}^{\prime}\cdot \underbar{p}),
\end{equation}
where $\underbar{q} = (q_1, q_2,\dots,  q_n)^T$, $\underbar{p}
= (p_1, p_2, \dots p_n)^T $ and $\xi = (q_{1}, p_{1}, \dots,
 q_{n}, p_{n})^{T}$. Thus, $W(\xi)$ is a
function of $2n$ real phase space variables  for an $n$-mode
quantum system.  The first order moments (sometimes called
displacement or mean) vector is given as
\begin{equation}
\overline{ \hat{\xi} } = \text{Tr}[\hat{\rho}\hat{\xi}].
\end{equation}
Similarly, the second order moments, which are best
represented in the form of a $2n \times 2n$,  real symmetric
matrix the called covariance matrix is given as
\begin{equation}\label{eq:cov}
V = (V_{ij})=\frac{1}{2}\langle \{\Delta \hat{\xi}_i,\Delta
\hat{\xi}_j\} \rangle,
\end{equation}
where $\Delta \hat{\xi}_i = \hat{\xi}_i-\langle \hat{\xi}_i
\rangle$, and $\{\,, \, \}$ denotes anticommutator. The uncertainty
principle obeyed by all quantum states can be
expressed easily in terms of covariance matrix as
\begin{equation}\label{eq:uncertainty}
V+\frac{i}{2}\Omega \geq 0 
\end{equation}

The class of Gaussian states, which is the focus of our
work is defined as a set of states with a Gaussian Wigner
distribution given as~\cite{weedbrook-rmp-2012}
\begin{equation}\label{eq:wignercovariance}
W(\xi) = \frac{\exp[-(1/2)(\xi-\overline{\hat{\xi}
})^TV^{-1}(\xi-\overline{ \hat{\xi} })]}{(2 \pi)^n
\sqrt{\text{det}V}},
\end{equation}
where $V$ is the covariance matrix and  $\overline{\xi} $
denotes the displacement of the Gaussian state.

As an example, the covariance matrix of single mode vacuum
state $\hat{\rho} = |0 \rangle\langle 0|$ is written as
\begin{equation}
V_{|0 \rangle}  = \frac{1}{2}\begin{pmatrix}
\langle \{ \Delta q, \Delta q\} \rangle & \langle \{ \Delta
q, \Delta p\} \rangle\\
\langle \{ \Delta p, \Delta q\} \rangle &
\langle \{ \Delta p, \Delta p\} \rangle
\end{pmatrix} =\frac{1}{2} \begin{pmatrix}
1 & 0 \\
0 & 1
\end{pmatrix}.
\end{equation}
Similarly, for a single mode thermal state,
\begin{equation}
\hat{\rho} = \sum_{n=0}^{\infty}\frac{\langle n \rangle ^n}
{(1+\langle n \rangle)^{n+1}}|n\rangle \langle n|,
\end{equation}
where $\langle n \rangle = 1/(\exp(\omega/k_B T)-1)$ is the average
number of photons in the thermal state, 
the covariance matrix is given by
\begin{equation}
V_{\text{th}}   =\frac{1}{2} \begin{pmatrix}
2 \langle n \rangle+1 & 0 \\
0 & 2 \langle n \rangle+1
\end{pmatrix}.
\end{equation}
We note that the vacuum state and the thermal state are
prominent examples of Gaussian states.

\par
\noindent{\bf Symplectic transformation\,:}
The linear homogeneous transformations described by real 
$2n \times 2n$ matrices $S$ act on the quadrature operators as $\hat{\xi}_i \rightarrow
\hat{\xi}_i^{\prime} = S_{ij}\hat{\xi}_{j} $. These matrices $S$
form a non-compact group called the symplectic group $Sp(2n, \, \mathcal{R})$
in $2n$ dimensions if they 
the following condition in order 
to preserve the  canonical commutation relation
Eq.~(\ref{eq:ccr}) and hence satisfy the condition:
\begin{equation}\label{symcond}
	\begin{aligned}
		Sp(2n, \, \mathcal{R})& = \{ S\,|\,S\Omega S^T = \Omega\}.
	\end{aligned}
\end{equation}
The symplectic transformation $S$ acts on the  Hilbert space of the
system via its infinite dimensional unitary representation
$\mathcal{U}(S)$, also known as metaplectic representation.
The elements of this metaplectic representations 
are generated by quadratic Hamiltonians.
Under a symplectic transformation $S$,  density operator,
mean and covariance matrix for any quantum state transform as
\begin{equation}\label{transformation}
	\rho \rightarrow \,\mathcal{U}(S) \rho
	\,\mathcal{U}(S)^{\dagger} \implies \overline{\hat{\xi}}\rightarrow S\overline{\hat{\xi}},\quad V\rightarrow SVS^T.
\end{equation}

In this paper we will be focusing on single mode Gaussian
states with diagonal covariance matrix with displacement
vector $\overline{\hat{\xi}}$ and covariance matrix $V$ given by
\begin{equation}\label{systemstate}
	\overline{\hat{\xi}}=  \begin{pmatrix}
		q_0 \\
		p_0 
	\end{pmatrix},\quad
	V=  \begin{pmatrix}
		(\Delta q)^2& 0 \\
		0 & (\Delta p)^2
	\end{pmatrix},
\end{equation}
where $q_0 = \langle \hat{q} \rangle$, $p_0 = \langle
\hat{p} \rangle$, $(\Delta q)^2 = \langle \hat{q}^2 \rangle
-(\langle \hat{q} \rangle)^2$, and $(\Delta p)^2 = \langle
\hat{p}^2 \rangle -(\langle \hat{p} \rangle)^2$.
The corresponding Gaussian Wigner distribution 
as per Eq.~(\ref{eq:wignercovariance}) is given as:
\begin{equation}\label{statewigner}
	W(q,p) = \frac{1}{2 \pi \Delta q \Delta
p}\exp\left[-\frac{(q-q_0)^2}{2 (\Delta
q)^2}-\frac{(p-p_0)^2}{2 (\Delta p)^2}\right].
\end{equation}
For analysis purposes in the later sections,  we shall use the
explicit form of the covariance matrix for  squeezed coherent
thermal state corresponding to temperature $T$ of a single mode 
system with frequency $\omega$,  which is given by
\begin{equation}\label{scts}
V= S(r)V_{\mathrm{th}}S(r)^T=\frac{1}{2}\begin{pmatrix}
(2 \langle n\rangle+1) e^{-2 r} & 0 \\
0 &  (2 \langle n\rangle+1) e^{2 r} \\
\end{pmatrix},
\end{equation}
where $S(r)$ is the single
mode squeezing transformation given by
\begin{equation}
S(r) = \begin{pmatrix}
e^{-r} & 0 \\
0 & e^{r}
\end{pmatrix}.
\end{equation}
This is the family of states that we will estimate using
different techniques in this paper.
\subsection{Measurement schemes}\label{schemes}
In this section, various measurement
schemes for state estimation, which we
intend to compare are described. A
Gaussian model for measurements is assumed, where any
measured quantity when repeatedly measured is fitted to a
Gaussian in the sense that we infer mean and variance from the
distribution. 
The variance of the probability distribution
signifies the
accuracy of the corresponding measurement scheme.
We assume that the
measurement apparatuses have infinite precision and the
variance in the outcomes for any measurement is due to the
inherent uncertainty in the observables.
\subsubsection{Homodyne measurement}
 
In homodyne measurement, we perform the measurement of
either quadrature $\hat{q}$ or the quadrature $\hat{p}$  on
the system~\cite{yuen-83,abbas-83,schumaker-1984,banaszek-pra-1997}. 
 The probability distribution function  $P(q)$
of obtaining the outcome `$q$' corresponding to the
measurement of the $\hat{q}$  quadrature on the state
given in Eq.~(\ref{systemstate}) can be evaluated  as
\begin{equation}
\begin{aligned}
P(q) =\int W(q,p) dp= \frac{1}{\sqrt{2 \pi (\Delta q)^2}} \exp
\left[-\frac{(q-q_0)^2}{2 (\Delta q)^2}\right].
\end{aligned}
\end{equation}
Therefore, the corresponding variance for the $\hat{q}$
quadrature  measurement is
\begin{equation}\label{homoq}
V^{\text{Hom}}(\hat{q})=(\Delta q)^2.
\end{equation}
 This variance is for infinite ensemble limit. Now if we have to
  estimate the $\hat{p}$ quadrature, we need to use a distinct
	ensemble that is prepared in the same way and that has not
	undergone the $\hat{q}$ quadrature measurement. 
  The probability distribution function $P(p)$  of
obtaining the outcome `$p$' corresponding to the measurement
of the $\hat{p}$ quadrature on the state~(\ref{systemstate})
can be evaluated as
\begin{equation}
\begin{aligned}
P(p) = \int W(q,p) dq = \frac{1}{\sqrt{2 \pi (\Delta
p)^2}}\exp\left[-\frac{(p-p_0)^2}{2 (\Delta p)^2}\right].
\end{aligned}
\end{equation}
Therefore, the corresponding variance for the $\hat{p}$
quadrature  measurement is
\begin{equation}\label{homop}
V^{\text{Hom}}(\hat{p})=(\Delta p)^2 .
\end{equation}
 
\subsubsection{Heterodyne measurement}
In heterodyne measurement, we jointly measure  the $\hat{q}$
quadrature and the $\hat{p}$  quadrature on the system~\cite{javan-62,read-1965,carleton-1968,gerhardt-1972,yuen-1980,
	yuen-1982,arthurs-1982,shapiro-1984, shapiro-1985,
	gardiner-1987, martens-pla-1991, raymer-apj-1994}.
Since the heterodyne measurement corresponds to the
projection onto coherent state basis
$E_{\alpha} = (2
\pi)^{-1}|\alpha\rangle \langle \alpha|$, the probability of
obtaining the outcomes `$q$' and `$p$' on the joint
measurement of the $q$ and $p$-quadratures on a system with
density operator $\rho$ can be written as
\begin{equation}
P(q,p) =\frac{1}{2\pi} \mathrm{Tr}[\hat{\rho}|\alpha\rangle
\langle \alpha |].
\end{equation}
For simplicity, we move to the phase space and evaluate 
the trace in the Wigner function description as follows:
\begin{equation}\label{tre}
P(q,p)=   \int_{\mathcal{R}^2} dq dp\, W_{\hat{\rho}}(q,p)
W_{|\alpha\rangle}(q,p).
\end{equation}
The probability
distribution function can then be calculated and is given
by:
\begin{equation}
P(q,p) = \frac{
\exp \left[-\frac{\displaystyle (q-q_0)^2}{1+2
(\displaystyle \Delta
q)^2}
-\frac{\displaystyle (p-p_0)^2}{\displaystyle (1+2 (\Delta
p)^2)}\right]
}
{\pi \sqrt{ \left(1+2(\Delta q)^2\right)\left(1+2(\Delta
p)^2\right)}}
\end{equation}
Therefore, the corresponding variance of the marginals
$P(q)$ and
$P(p)$ of the probability distribution  $P(q,p)$ is
\begin{equation}\label{heteropq}
V^{\text{Het}}(\hat{q})= \frac{1}{2}+(\Delta q)^2,\quad
\text{and} 
\quad V^{\text{Het}}(\hat{p})=\frac{1}{2}+(\Delta p)^2 .
\end{equation}
We note that the vacuum noise (equal to $1/2$) is added to
the variance of both the marginals $P(q)$ and $P(p)$ of the
probability distribution  $P(q,p)$.

\subsubsection{Sequential measurement}\label{sec:sequential}
\begin{figure}[h!]
\centering
\scalebox{1.0}{\includegraphics{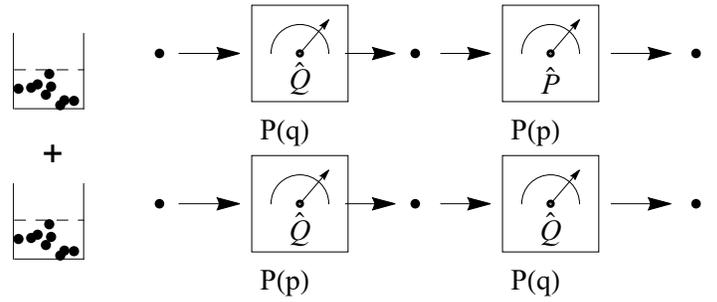}}
\caption{Schematic representation of the sequential
measurement scheme. The whole ensemble is divided in two
halves.  On the first half, the sequential measurement of
the $\hat{Q}$-quadrature of the meter, which renders
information about the $\hat{q}$-quadrature of the system,
is followed by a homodyne measurement of the
$\hat{p}$-quadrature of the system. Similarly, on the second
half, the sequential measurement of the $\hat{Q}$-quadrature
of the meter, which renders information about the
$\hat{p}$-quadrature  of the system, is followed by a
homodyne measurement of the $\hat{q}$-quadrature of the
system.}
\label{schemesequential}
\end{figure}

In the sequential measurement scheme,  the measurement of
one quadrature is followed by the measurement of its
conjugate quadrature~\cite{jukka-pra-2009,lorenzo-prl-2013,da-jpa-2017}. To carry out the first measurement,
we use the von-Neumann measurement model, where we couple
the system with a meter. The state of the system is inferred
through the readings of the meter. The second measurement,
\ie, the measurement of the conjugate observable, is a
homodyne measurement.  To remove the biasedness in the order
of the measurements, we divide the ensemble in two
halves~\cite{da-jpa-2017}.  On the first half,  we measure
the $\hat{Q}$-quadrature of the meter weakly, which renders
information about the $\hat{q}$-quadrature of the system.
This weak measurement disturbs the state a little and does
not lead to the complete collapse of the wave
function, and therefore, the state can be reused for
measurement.  This is followed by a homodyne measurement of
the $\hat{p}$-quadrature on the system, as shown in
Fig.~\ref{schemesequential}.  Similarly, on the second half,
the weak measurement of $\hat{Q}$-quadrature of the meter
renders information about the $\hat{p}$-quadrature of the
system, which is followed by a homodyne measurement of the
$\hat{q}$-quadrature of the system.

We now describe the scheme
in detail.  While the
system is represented by the quadrature operators $\hat{q}$
and $\hat{p}$, we consider the apparatus also to be a one
mode CV system representing a meter with quadrature
operators $\hat{Q}_1$ and $\hat{P}_1$.  The corresponding
phase space is four-dimensional and can be represented by
four variables, which can be arranged in a column vector
form as $\xi = (q,p,Q_1,P_1)^T$. We assume that the system
is in a squeezed coherent thermal state and the meter is in
a squeezed vacuum state, and thus they satisfy the following
uncertainty relations:
\begin{equation}
\Delta q \Delta p \geq 1/2,\quad \Delta Q_1 \Delta P_1 = 1/2.
\end{equation}
Since the system and the meter are in Gaussian states, the 
system-meter state  can be specified by the following
displacement vector and covariance matrix:
\begin{equation}\label{disi}
\overline{\hat{\xi}}= \begin{pmatrix}
\langle \hat{q}\rangle =  q_0 \! \\
\langle \hat{p}\rangle =  p_0 \!\\
\langle \hat{Q}_1\rangle = \! 0 \! \\
\langle \hat{P}_1\rangle = \! 0\! \\
\end{pmatrix}\!\!,\, 	V = \begin{pmatrix}
(\Delta q)^2 \!\!& 0 & 0 & 0 \\
0 & (\Delta p)^2\!\! & 0 & 0 \\
0 & 0 & (\Delta Q_1)^2\!\! & 0 \\
0 & 0 & 0 & (\Delta P_1)^2 \!\!\\
\end{pmatrix}.
\end{equation}
Now we consider the interaction Hamiltonian of the form
\begin{equation}\label{intseq}
\hat{H}(t) = \delta(t-t_1)\hat{q}\hat{P_1},
\end{equation}
which entangles the system and the meter. The unitary operator
acting on the joint system-meter Hilbert space for $t>t_1$ is given by
\begin{equation}
\mathcal{U}(\hat{H}(t)) =\mathrm{e}^{-i\int \hat{H}(t) \,dt}
=\mathrm{e}^ { -i\hat{q}\hat{P_1} }.
\end{equation}
The corresponding symplectic transformation acting on the 
quadrature operators  $\hat{\xi} =
(\hat{q},\hat{p},\hat{Q}_1,\hat{P}_1)^T$
is given by (see Appendix~\ref{symplecticcalculation})
\begin{equation}
S = \begin{pmatrix}
1 & 0 & 0 & 0 \\
0 & 1 & 0 & -1 \\
1 & 0 & 1 & 0 \\
0 & 0 & 0 & 1 \\
\end{pmatrix}.
\end{equation}
The above symplectic transformation is an element of the
real symplectic group $Sp(4,\, \mathcal{R})$
and satisfies the symplectic condition~(\ref{symcond}).
As a result of the above transformation,
the final displacement vector and covariance matrix
$V$~(\ref{disi}) 
can be written as follows according to
Eq.~(\ref{transformation}):
\begin{equation}\label{disf}
\overline{\hat{\xi}^{'}} = \begin{pmatrix}
\langle \hat{q}\rangle =  q_0  \\
\langle \hat{p}\rangle =  p_0 \\
\langle \hat{Q}_1\rangle = q_0  \\
\langle \hat{P}_1\rangle =  0\\
\end{pmatrix}, \quad \text{and}
\end{equation}
\begin{equation}\label{covf}
V^{'} = \begin{psmallmatrix}
(\Delta q)^2 & 0 & (\Delta q)^2 & 0 \\
0 & (\Delta p)^2+(\Delta P_1)^2 & 0 & -(\Delta P_1)^2 \\
(\Delta q)^2 & 0 & (\Delta q)^2 +(\Delta Q_1)^2& 0 \\
0 & -(\Delta P_1)^2 & 0 & (\Delta P_1)^2 \\
\end{psmallmatrix}.
\end{equation}
One can easily find the transformed Wigner distribution  of
the system-meter using Eq.~(\ref{eq:wignercovariance}),
which is specified by the displacement vector~(\ref{disf})
and the
covariance matrix~(\ref{covf}). The Wigner distribution of
the reduced state of the meter can be evaluated by
integrating the system-meter Wigner distribution over the
system variables $q$ and $p$. The displacement vector and
the covariance matrix of the reduced state can be readily
evaluated using the Wigner function of the reduced state. An
alternative to this approach is to work at the covariance
matrix level.  The displacement vector and the covariance
matrix of the reduced state of the meter can be obtained by
ignoring the matrix elements corresponding to the system
mode. This can be easily seen through the Wigner
characteristic function of a Gaussian
state~\cite{adesso-2014,olivares-2012}.  Thus,  the
displacement vector and the covariance matrix  of the
reduced state of the meter are
\begin{equation}\label{covarsequential1}
\overline{\hat{\xi}_{\mathrm{M}}^{'}} = \begin{pmatrix}
\langle \hat{Q}_1\rangle = q_0  \\
\langle \hat{P}_1\rangle =  0\\
\end{pmatrix}, \quad V^{'}_{\mathrm{M}} = \begin{pmatrix}
(\Delta q)^2 +(\Delta Q_1)^2& 0 \\
0 & (\Delta P_1)^2 \\
\end{pmatrix}.
\end{equation}
The corresponding Wigner function for the reduced state of
the meter can be written
using Eq.~(\ref{eq:wignercovariance}) as
\begin{equation}
\begin{aligned}
W(Q_1,P_1)& = \frac{1}
{2\pi \Delta P_1 \sqrt{ (\Delta q)^2+(\Delta Q_1)^2}}\\
&\times	\exp \left[-\frac{(Q_1-q_0)^2}
{2((\Delta q)^2+(\Delta Q_1)^2)}-\frac{P_1^2}{2 (\Delta
P_1)^2}\right].
\end{aligned}
\end{equation}
The probability density to obtain the outcome $Q_1$ after a
measurement of the $\hat{Q}_1$-quadrature on the meter is
\begin{equation}
\begin{aligned}
P(Q_1) &= \frac{1}
{  \sqrt{ 2\pi((\Delta q)^2+(\Delta Q_1)^2)}}\\
&  \times	\exp \left[-\frac{(Q_1-q_0)^2}
{2((\Delta q)^2+(\Delta Q_1)^2)}\right].
\end{aligned}
\end{equation}
Clearly, the variance of the probability distribution is
\begin{equation}\label{sm1q}
V_1^{\mathrm{SM}}(\hat{q}) =  (\Delta q)^2+(\Delta Q_1)^2,
\end{equation}
which we could have directly written from Eq.~(\ref{covf})
as the element corresponding to the variance of $Q_1$.
Similarly,  
the displacement vector and the covariance matrix
of the reduced state of the system are given by
\begin{equation}\label{covarsequential2}
\overline{\xi_{\mathrm{S}}^{'} }= \begin{pmatrix}
\langle \hat{Q}_1\rangle = q_0  \\
\langle \hat{P}_1\rangle =  p_0\\
\end{pmatrix}, \quad	V^{'}_{\mathrm{S}} = \begin{pmatrix}
(\Delta q)^2 & 0 \\
0 & (\Delta p)^2+(\Delta P_1)^2 \\
\end{pmatrix}.
\end{equation}
The variance of the probability distribution corresponding
to the homodyne
measurement of the $\hat{p}$-quadrature is given by
\begin{equation}\label{sm1p}
V_1^{\mathrm{SM}}(\hat{p}) =  (\Delta p)^2+(\Delta P_1)^2.
\end{equation}

We now discuss the weak measurement of the
$\hat{p}$-quadrature followed by a homodyne measurement of
the $\hat{q}$-quadrature.  We again consider the initial
state of the joint system-meter state being represented by
the displacement vector and the covariance matrix as given
in Eq.~(\ref{disi}).  We consider the interaction
Hamiltonian of the form
\begin{equation}
\hat{H}(t) = \delta(t-t_1)\hat{p}\hat{Q_1},
\end{equation}
which is the generator of the following symplectic
transformation:
\begin{equation}
S = \begin{pmatrix}
1 & 0 & 0 & 1 \\
0 & 1 & 0 & 0 \\
0 & 1 & 1 & 0 \\
0 & 0 & 0 & 1 \\
\end{pmatrix}.
\end{equation}
The above symplectic transformation is also an element of
the real symplectic group $Sp(4,\, \mathcal{R})$
and satisfies the symplectic condition~(\ref{symcond}).
The final system-meter state after the action of the above 
symplectic transformation can be specified by the following
displacement vector and covariance matrix:
\begin{equation}\label{disreverse}
\overline{\hat{\xi}^{'}}   = \begin{pmatrix}
\langle \hat{q}\rangle =  q_0  \\
\langle \hat{p}\rangle =  p_0 \\
\langle \hat{Q}_1\rangle = p_0  \\
\langle \hat{P}_1\rangle =  0\\
\end{pmatrix}, \quad \text{and}
\end{equation}
\begin{equation}\label{covreverse}
V^{'} = \begin{psmallmatrix}
(\Delta q)^2 +(\Delta P_1)^2& 0 & 0 & (\Delta P_1)^2 \\
0 & (\Delta p)^2 & (\Delta p)^2 & 0 \\
0 & (\Delta p)^2 & (\Delta p)^2+(\Delta Q_1)^2 & 0 \\
(\Delta P_1)^2 & 0 & 0 & (\Delta P_1)^2 \\
\end{psmallmatrix}.
\end{equation}
The displacement vector~(\ref{disreverse}) shows that the
mean of the $\hat{Q}_1$-quadrature for the meter is $p_0$.
Thus, the measurement of the $\hat{Q}_1$-quadrature of the
meter yields information about the $\hat{p}$ quadrature of
the system. We can directly write the variance of the
probability distributions corresponding to the sequential
measurement of the $\hat{Q}_1$-quadrature of the meter
followed by a homodyne measurement of the
$\hat{q}$-quadrature from Eq.~(\ref{disreverse}) as
\begin{equation}
\begin{aligned}\label{sm2pq}
V_2^{\mathrm{SM}}(\hat{p}) =  & (\Delta q)^2 +(\Delta
P_1)^2,\quad \mathrm{and} \\
\quad V_2^{\mathrm{SM}}(\hat{q}) =&  (\Delta p)^2+(\Delta
Q_1)^2.
\end{aligned}
\end{equation}
\subsubsection{Arthurs-Kelly measurement scheme}
\begin{figure}[H]
\centering
\scalebox{1.0}{\includegraphics{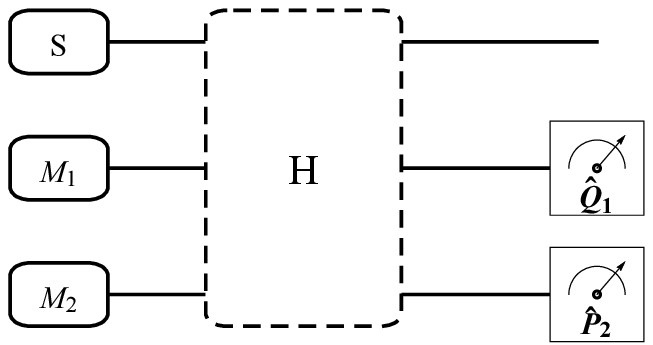}}
\caption{Schematic representation of the Arthurs-Kelly
scheme.  The system is labeled by $S$, while the two meters
are labeled by $M_1$ and $M_2$. $H$ represents the
interaction Hamiltonian.  Measurement of
$\hat{Q}_1$-quadrature on meter $M_1$ and  measurement of
$\hat{P}_2$-quadrature on meter $M_2$ yield information
about the $\hat{q}$-quadrature and the $\hat{p}$-quadrature
of the system, respectively.}
\label{akscheme}
\end{figure}
Arthurs-Kelly proposed a scheme by extending the von Neumann
model, which enables us to simultaneously
measure conjugate
quadratures $\hat{q}$ and $\hat{p}$~\cite{arthurkelly-1964}. 
 To this end, two
meters, one for each quadrature measurement, are introduced,
as shown in Fig.~\ref{akscheme}. 
We represent the system and two meters using three pairs of
Hermitian quadrature operators arranged in column vector as
$\hat{\xi} =
(\hat{q},\hat{p},\hat{Q}_1,\hat{P}_1,\hat{Q}_2,\hat{P}_2)^T$,
where $(\hat{q},\hat{p})$ corresponds to the system, and
$(\hat{Q}_1,\hat{P}_1)$ and $(\hat{Q}_2,\hat{P}_2)$
correspond to the two meters. We assume the system to be  in a
squeezed coherent thermal state and the meters  to be  in a
squeezed vacuum state, and thus they satisfy the following
uncertainty relations:
\begin{equation}
\Delta q \Delta p \geq 1/2,\quad \Delta Q_1 
\Delta P_1 = 1/2 ,\quad \Delta Q_2 \Delta P_2 = 1/2.
\end{equation}
We analyze our joint system in a six-dimensional phase 
space represented by six variables, which can be arranged in 
a column vector form as  $\xi = (q,p,Q_1,P_1,Q_2,P_2)^T$.
We represent the system-meters state by the displacement vector 
$\overline{\hat{\xi}}$ and covariance matrix $V$ as
\begin{widetext}
\begin{equation}  
\overline{\hat{\xi}}=  \begin{pmatrix}
\langle \hat{q}\rangle = q_0 \\
\langle \hat{p}\rangle = p_0 \\
\langle \hat{Q_1}\rangle = 0 \\
\langle \hat{P_1}\rangle = 0  \\
\langle \hat{Q_2}\rangle = 0  \\
\langle \hat{P_2}\rangle = 0 \\
\end{pmatrix}, \quad
V=\begin{tikzpicture}[baseline=(m-3-1.base),
every left delimiter/.style={xshift=.75em},
every right delimiter/.style={xshift=-.75em},
]
\matrix [matrix of math nodes,left delimiter=(,right delimiter=),row sep=0.01cm,column sep=0.01cm] (m) {
	(\Delta q)^2&0&0&0&0&0 \\
	0&(\Delta p)^2&0&0&0&0 \\
	0&0&(\Delta Q_1)^2&0&0&0 \\
	0&0&0&(\Delta P_1)^2&0&0 \\
	0&0&0&0&(\Delta Q_2)^2&0 \\
	0&0&0&0&0&(\Delta P_2)^2 \\};
		
\draw[solid] ($0.5*(m-1-2.north east)+0.5*(m-1-3.north west)$) --
($0.5*(m-6-3.south east)+0.5*(m-6-2.south west)$);
\draw[solid] ($0.5*(m-1-4.north east)+0.5*(m-1-5.north west)$) --
($0.5*(m-6-5.south east)+0.5*(m-6-4.south west)$);
		
\draw[solid] ($0.5*(m-2-1.south west)+0.5*(m-3-1.north west)$) --
($0.5*(m-2-6.south east)+0.5*(m-3-6.north east)$);
	
\draw[solid] ($0.5*(m-4-1.south west)+0.5*(m-5-1.north west)$) --
($0.5*(m-4-6.south east)+0.5*(m-5-6.north east)$);
\node[above=1pt of m-1-1] (top-1) {$q$};
\node[above=1pt of m-1-2] (top-2) {$p$};
\node[above=1pt of m-1-3] (top-3) {$Q_1$};
\node[above=1pt of m-1-4] (top-4) {$P_1$};
\node[above=1pt of m-1-5] (top-5) {$Q_2$};
\node[above=1pt of m-1-6] (top-6) {$P_2$};

\coordinate[above=3pt of top-1] (aux);
\begin{scope}[nodes={text depth=0.25ex}]
\draw[thick,decorate,decoration=brace] (top-1.west|-aux) -- (top-2.east|-aux)
node[midway,above=0.1ex] {System};
\draw[thick,decorate,decoration=brace] (top-3.west|-aux) -- (top-4.east|-aux)
node[midway,above=0.1ex] {Meter 1};
	\draw[thick,decorate,decoration=brace] (top-5.west|-aux) -- (top-6.east|-aux)
	node[midway,above=0.1ex] {Meter 2};
	\end{scope}

\node[right=18pt of m-1-6] (left-1) {$q$};
\node[right=18pt of m-2-6] (left-1) {$p$};
\node[right=18pt of m-3-6] (left-1) {$Q_1$};
\node[right=18pt of m-4-6] (left-1) {$P_1$};
\node[right=18pt of m-5-6] (left-1) {$Q_2$};
\node[right=2pt of m-6-6] (left-1) {$P_2$};
\end{tikzpicture}.
\end{equation}
\end{widetext}
The interaction Hamiltonian through which we intend to
measure both the system quadratures by coupling them to
different meters is considered to be of the form
\begin{equation}
H =\delta(t-t_1) (\hat{q} \hat{P}_1-\hat{p}\hat{ Q}_2),
\end{equation}
which entangles the system with both the meters.  The
corresponding symplectic transformation acting on the
quadrature operators $\hat{\xi}$ is given by
\begin{equation}  
S=\begin{tikzpicture}[baseline=(m-3-1.base),
every left delimiter/.style={xshift=.75em},
every right delimiter/.style={xshift=-.75em},
]
\matrix [matrix of math nodes,left delimiter=(,right
delimiter=),row sep=0.01cm,column sep=0.01cm] (m) {
	1 & 0 & 0 & 0 & -1 & 0 \\
	0 & 1 & 0 & -1 & 0 & 0 \\
	1 & 0 & 1 & 0 &  -\frac{1}{2} &0 \\
	0 & 0 & 0 & 1 & 0 & 0 \\
	0 & 0 & 0 &0  & 1 & 0 \\
	0 & 1 & 0 & -\frac{1}{2} & 0 & 1 \\};
\draw[solid] ($0.5*(m-1-2.north east)+0.5*(m-1-3.north west)$) --
($0.5*(m-6-3.south east)+0.5*(m-6-2.south west)$);
\draw[solid] ($0.5*(m-1-4.north east)+0.5*(m-1-5.north west)$) --
($0.5*(m-6-5.south east)+0.5*(m-6-4.south west)$);
\draw[solid] ($0.5*(m-2-1.south west)+0.5*(m-3-1.north west)$) --
($0.5*(m-2-6.south east)+0.5*(m-3-6.north east)$);
\draw[solid] ($0.5*(m-4-1.south west)+0.5*(m-5-1.north west)$) --
($0.5*(m-4-6.south east)+0.5*(m-5-6.north east)$);
\node[above=1pt of m-1-1] (top-1) {$q$};
\node[above=1pt of m-1-2] (top-2) {$p$};
\node[above=1pt of m-1-3] (top-3) {$Q_1$};
\node[above=1pt of m-1-4] (top-4) {$P_1$};
\node[above=1pt of m-1-5] (top-5) {$Q_2$};
\node[above=1pt of m-1-6] (top-6) {$P_2$};
\node[right=2pt of m-1-6] (left-1) {$q$};
\node[right=2pt of m-2-6] (left-1) {$p$};
\node[right=2pt of m-3-6] (left-1) {$Q_1$};
\node[right=2pt of m-4-6] (left-1) {$P_1$};
\node[right=2pt of m-5-6] (left-1) {$Q_2$};
\node[right=2pt of m-6-6] (left-1) {$P_2$};
\end{tikzpicture}.
\end{equation}
The above symplectic transformation is an element of the
real symplectic group $Sp(6,\, \mathcal{R})$ and satisfies
the symplectic condition~(\ref{symcond}). The displacement
 vector and the covariance matrix of the transformed joint
system-meters state is explicitly written in
Eqs.~(\ref{appendixdis}) and
(\ref{appendixcov}) of Appendix~\ref{appendixAK}.
The displacement vector and the covariance matrix
of the reduced state of the system can be readily 
written using Eqs.~(\ref{appendixdis}) and
(\ref{appendixcov}) as
\begin{equation}
\overline{\xi_{\text{S}}^{'} }= \begin{psmallmatrix}
\langle \hat{Q}_1\rangle = q_0  \\
\langle \hat{P}_1\rangle =  p_0\\
\end{psmallmatrix}, \quad 	V^{'}_{\text{S}} = \begin{psmallmatrix}
(\Delta q)^2+ (\Delta Q_2)^2 & 0\\
0 &(\Delta p)^2+ (\Delta P_1)^2 \\
\end{psmallmatrix}.
\end{equation}
Similarly, the displacement vector and the covariance matrix
of  meter $1$ can be written as
\begin{equation}
\overline{\xi_{\text{M}_1}^{'} }= \begin{psmallmatrix}
\langle \hat{Q}_1\rangle = q_0  \\
\langle \hat{P}_1\rangle =  0\\
\end{psmallmatrix}, \quad V^{'}_{\text{M}_1} =
\begin{psmallmatrix}
(\Delta q)^2+(\Delta Q_1)^2+ \frac{(\Delta Q_2)^2}{4} & 0\\
0 & (\Delta P_1)^2 \\
\end{psmallmatrix}.
\end{equation}
Finally, meter $2$ is represented by the following
displacement vector and covariance matrix:
\begin{equation}
\overline{\xi_{\text{M}_2}^{'} }= \begin{psmallmatrix}
\langle \hat{Q}_1\rangle = 0 \\
\langle \hat{P}_1\rangle =  p_0\\
\end{psmallmatrix}, \quad 	V^{'}_{\text{M}_2} = \begin{psmallmatrix}
(\Delta Q_2)^2 & 0\\
0& (\Delta p)^2+\frac{(\Delta P_1)^2}{4}+(\Delta P_2)^2\\
\end{psmallmatrix}.
\end{equation}
The variance of the probability distribution for the
measurement of the $\hat{Q}_1$-quadrature on the meter $1$
and $\hat{P}_2$-quadrature on meter $2$ can be directly
written  from  Eq.~(\ref{appendixcov}) as
\begin{equation}\label{akqp}
\begin{aligned}
V^{\mathrm{AK}}(\hat{q}) = & (\Delta q)^2+(\Delta Q_1)^2+
\frac{(\Delta Q_2)^2}{4},\\
V^{\mathrm{AK}}(\hat{p}) = & (\Delta p)^2+\frac{(\Delta
P_1)^2}{4}+(\Delta P_2)^2.
\end{aligned}
\end{equation}
\section{Results}\label{results}
In this section we turn to the examination of
the performance of
measurement schemes described in
Section~\ref{background} in the estimation of the Wigner
distribution of an ensemble with a fixed number $N$ of
identically prepared Gaussian states. To this end, 
we define a distance measure $d_1$ for
the accuracy estimation of the mean of the Gaussian state as
\begin{equation}
d_1 = \langle (q^{A}-q^{M})^2\rangle+\langle (p^{A}-
p^{M})^2\rangle,
\label{fid1}
\end{equation}
where $q^A$ and $p^A$ are the actual values of the mean of
the $\hat{q}$ and $\hat{p}$ quadratures of the Gaussian
state and are thus fixed, whereas $q^M$ and $p^M$ are the
measured values of the $\hat{q}$ and $\hat{p}$ quadratures
of the Gaussian state.   While $q^A$ and $p^A$ are the same 
for each copy of the ensemble, the values $q^M$ and $p^M$ 
obtained by measuring different copies of the ensemble
 can be different. The magnitude of the
distance measure $d_1$ signifies how well the mean $(q_0,
p_0)$ of the Gaussian state  has been estimated. We define
another distance measure  $d_2$ for the accuracy estimation
of the variance of the
Gaussian state as
\begin{equation}
d_2= \langle (V_q^{A}- V_q^{M})^2\rangle+\langle (V_p^{A}-
V_p^{M})^2\rangle,
\label{fid2}
\end{equation}
where $V_q^{A}$ and  $V_p^{A}$ are the actual values of
the variance of  $\hat{q}$ and $\hat{p}$ quadratures, while 
$V_q^{M}$ and  $V_p^{M}$ are the measured values of the 
variance of  $\hat{q}$ and $\hat{p}$ quadratures. Here 
$d_2$ signifies how well the variance $(\Delta q)^2$ 
and $(\Delta p)^2$ has been estimated. In the case of perfect 
estimation, both the distance measures $d_1$ and $d_2$ should 
approach zero. 

\subsection{Analytical expressions of distance measure
$d_1$} Now we evaluate the distance measure $d_1$ for
various measurement schemes, which are employed for the estimation
of Gaussian states.

\noindent
{\bf Homodyne scheme:}
To estimate the state using the
homodyne measurement, we divide the ensemble in two halves.
On the first half of the ensemble, the $\hat{q}$-quadrature
is measured, while on the other half of the ensemble, the
$\hat{p}$-quadrature is measured. Thus, we can write the
distance measure $d_1$ for the homodyne measurement using
Eqs.~\eqref{homoq} and~\eqref{homop} as
\begin{equation}\label{vhomo}
d_1^{\mathrm{Hom}} = \frac{(\Delta q)^2}{N/2}+\frac{(\Delta p)^2}{N/2}.
\end{equation}
Here we have used the fact that the probability distribution
involved in the homodyne measurement is Gaussian, and the
 sample variance for a Gaussian (normal) distribution 
$\mathcal{N}(\mu, \sigma)$ with mean $\mu$ and variance
$\sigma^2$ for a sample of size $N$ 
is given by $\sigma^2/N$.

\noindent
{\bf Heterodyne scheme:}
The distance measure $d_1$ for the heterodyne measurement
can be calculated 
using Eq.~\eqref{heteropq} and is given  as
\begin{equation}\label{vhetero}
d_1^{\mathrm{Het}} = \frac{(\Delta q)^2+1/2}{N}+\frac{(\Delta p)^2+1/2}{N}.
\end{equation}
We can show  analytically from Eqs.~(\ref{vhomo}) and
(\ref{vhetero}) that $d_1^{\mathrm{Hom}} \geq d_1^{\mathrm{Het}}$,
where the equality sign only holds for a coherent state
ensemble.  Therefore, the homodyne measurement and the
heterodyne measurement perform the same for a coherent state
ensemble, whereas the heterodyne measurement outperforms the
homodyne measurement for a squeezed state ensemble as
far as the mean estimation is concerned.

\noindent
{\bf Sequential measurement scheme:}
For the sequential measurement scheme, we again divide the
ensemble in two halves and perform measurement according to
the procedure described in Sec.~\ref{sec:sequential}. 
In this case, the
expression of the distance measure $d_1$ turn out to be
\begin{equation}
d_1^{\mathrm{SM}} = \left\langle \left(q^{A}-
\frac{q^{M}_1+q^{M}_2}{2}\right)^2 \right\rangle+\left
\langle\left (p^{A}-
\frac{p^{M}_1+p^{M}_2}{2}\right)^2\right\rangle,
\label{fidsequential1}
\end{equation}
which can be re-written as follows using
Eqs.~\eqref{sm1q}, \eqref{sm1p}, and~\eqref{sm2pq}:
\begin{equation}
d_1^{\mathrm{SM}} = \frac{(\Delta q)^2+(\Delta p)^2
+(\Delta Q_1)^2+(\Delta P_1)^2}{N}.
\end{equation}
It can be seen from the above equation that the optimal
performance in the mean estimation for the sequential
measurement scheme corresponds to $\Delta Q_1=\Delta P_1=
1/\sqrt{2}$.  Further, at the optimal conditions,
$d_1^{\mathrm{SM}}=d_1^{\mathrm{Het}}$.  

\noindent
{\bf Arthurs-Kelly Scheme :} 
For the
Arthurs-Kelly scheme, we    
write the expression of distance measure $d_1$ 

 using Eq.~\eqref{akqp} as
\begin{equation}
\begin{aligned}
d_1^{\text{AK}}=&\frac{(\Delta q)^2+(\Delta Q_1)^2+ \frac{(\Delta Q_2)^2}{4}}{N}\\
&+\frac{(\Delta p)^2+\frac{(\Delta P_1)^2}{4}+(\Delta P_2)^2}{N}.
\end{aligned}
\end{equation}
For the Arthurs-Kelly scheme, the optimal performance in the
mean estimation corresponds to 
\begin{equation}\label{optimal}
	\Delta Q_1 =1/2, \quad\Delta P_2 = 1/2,
\end{equation}
and at the optimal conditions, $d_1^{\mathrm{AK}}=d_1^{\mathrm{Het}}$.
This means that the optimal performance in the mean estimation of the sequential measurement
requires only classical resources, \ie, the meter should be 
prepared in a coherent state, while the Arthurs-Kelly scheme requires 
nonclassical resources, \ie, the meters should be prepared in a squeezed state.

Now we illustrate  the dependence of the distance measure $d_1$ on the
initial width of the meter $\Delta Q_1$, the squeezing parameter $r$, and the
average number of photons $\langle n\rangle$
graphically. We have considered an ensemble
of size $N=20$ in all different plots in this article.

\begin{figure}[h!]
\centering
\scalebox{1.0}{\includegraphics{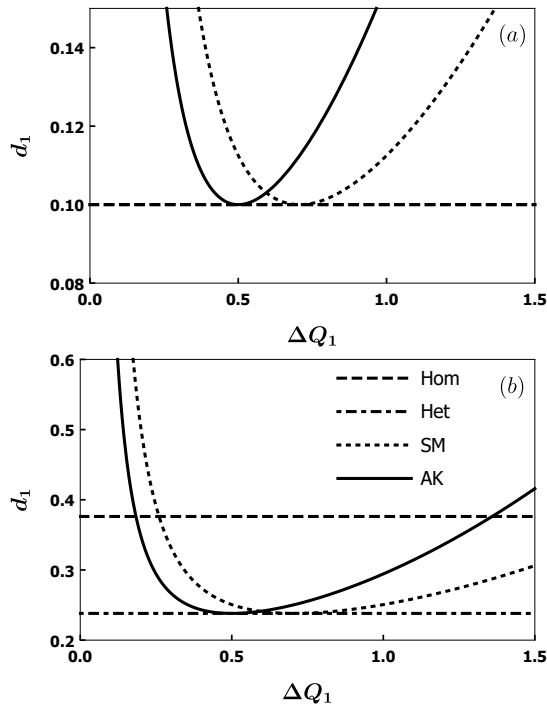}}
\caption{ Both the plots show the distance measure $d_1$ as
a function of the initial width of the meter $\Delta
Q_1$ for an ensemble of size $N=20$.  Additionally, we have
taken $\Delta P_2 = 1/2$ in all the graphs for the
Arthurs-Kelly scheme, which is the condition for the optimal
performance~\eqref{optimal}.  $(a)$ The ensemble consists of
identically prepared coherent states. The  homodyne
measurement and the heterodyne measurement perform equally
in this case. $(b)$ The ensemble consists of identically
prepared squeezed coherent states with squeezing parameter
$r=1$. }
\label{variance}
\end{figure}
We show the plot of the distance measure $d_1$ as a
function of the initial width of the meter $\Delta Q_1$ for
a coherent state ensemble in Fig.~\ref{variance}$(a)$. 
The results  show that the
homodyne measurement and the heterodyne measurement perform
the same, and the optimal performance of the Arthurs-Kelly
scheme and the sequential measurement schemes are equal to
that of the homodyne measurement and the heterodyne
measurement. Similarly, Fig.~\ref{variance}$(b)$ shows the
plot of distance measure $d_1$ as a function of the initial width of
the meter $\Delta Q_1$ for a squeezed coherent state
ensemble with squeezing parameter $r=1$. In this case,  the
heterodyne measurement outperforms the homodyne measurement.
Further, an increase or a decrease in the size of the
ensemble changes only the magnitude of the distance measure, while
the performance trend of the various measurement schemes
remain the same. It should be noted that these conclusions
about the relative performances of the various measurement 
schemes are based on the mean estimation efficacy.

\begin{figure}[h!]
\centering
\scalebox{1.0}{\includegraphics{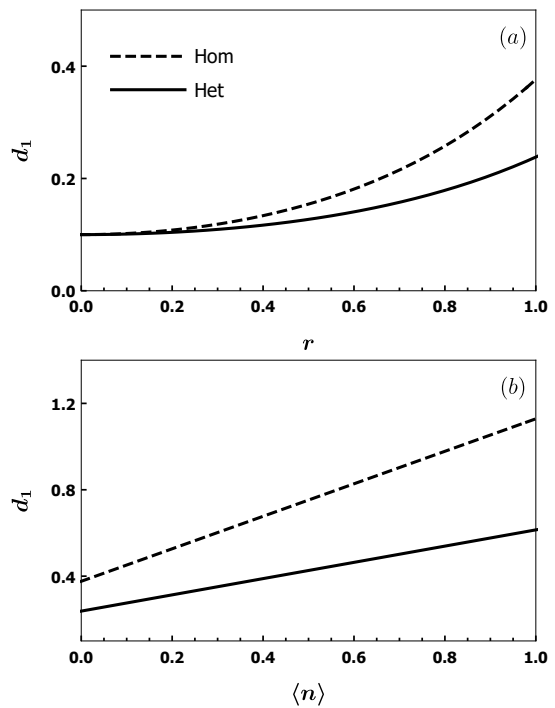}}
\caption{ $(a)$ The distance measure $d_1$ as a 
function of the squeezing parameter $r$. Here average number 
of photon is $\langle n \rangle=0$. $(b)$  The distance measure $d_1$ as a
function of the  average photon number $\langle n \rangle$.  
Here squeezing parameter  has been taken as $r=1$. An ensemble
of size $N=20$ has been considered for both the plots.  }
\label{vcomp}
\end{figure}
We plot the distance measure $d_1$ as a function of the squeezing
parameter $r$ in Fig.~\ref{vcomp}$(a)$. 
The results show
that both the homodyne measurement and the heterodyne
measurement estimate the mean of the Gaussian state with the
same distance measure for a coherent state ensemble $(r=0)$, while
for a squeezed coherent state ensemble $(r>0)$, the
heterodyne measurement outperforms the homodyne measurement.
The plot of the distance measure $d_1$ as a function of the average
number of photons $\langle n\rangle$ is shown in
Fig.~\ref{vcomp}$(b)$. The results  show that the distance measure
of the mean estimation increases, \ie, the estimation
efficiency decreases, for a thermal state $(\langle n\rangle
> 0)$ ensemble as compared to a pure state $(\langle
n\rangle =0)$ ensemble.
\subsection{Analytical expressions of  distance measure $d_2$}
We now proceed to derive expressions of distance measure 
$d_2$~(\ref{fid2}) for all the measurement schemes.

\noindent
{\bf Homodyne  Scheme :}
For the homodyne measurement, the expression of the distance
measure $d_2$ evaluates to
\begin{equation}
\begin{aligned}
d_2^{\mathrm{Hom}} = \frac{2(\Delta
q)^4}{N/2}+\frac{2(\Delta p)^4}{N/2}.
\end{aligned}
\end{equation}
Here we have used the fact that the variance of the sample
variance for a Gaussian (normal) distribution
$\mathcal{N}(\mu, \sigma)$ with mean $\mu$ and variance
$\sigma^2$ for a sample of size $N$ is given by
2$\sigma^4/N$.

\noindent
{\bf Heterodyne  Scheme :}
The expression of the distance
measure $d_2$ for the heterodyne measurement  evaluates to
\begin{equation}
\begin{aligned}
d_2^{\mathrm{Het}} = \frac{2 \left((\Delta
q)^2+1/2\right)^2}{N}+\frac{2\left((\Delta
p)^2+1/2\right)^2}{N}.
\end{aligned}
\end{equation}

\noindent
{\bf Sequential measurement  Scheme :}
For the sequential measurement scheme, the expression of the
distance measure  $d_2$ can be written in an analogous way as
Eq.~(\ref{fidsequential1}). The final expression of the
distance measure in this case evaluates to
\begin{equation}
\begin{aligned}
d_2^{\text{SM}}=& \frac{2}{N}\frac{}{}\bigg[(\Delta
p)^4+(\Delta q)^2 \left((\Delta Q_1)^2+(\Delta
q)^2\right)^2\\
&+(\Delta P_1)^2 \left((\Delta q)^2+(\Delta p)^2+(\Delta P_1)^2\right)^2\\
&+(\Delta Q_1)^2 \left((\Delta Q_1)^2+(\Delta p)^2\right)^2\bigg].
\end{aligned}
\end{equation}

\noindent
{\bf Arthurs-Kelly  Scheme :}
Similarly, the distance measure  $d_2$  for the Arthurs-Kelly scheme
can be calculated and turns out to
be:
\begin{equation}
\begin{aligned}
d_2^{\text{AK}}=&\frac{2\left((\Delta q)^2+(\Delta Q_1)^2+
\frac{(\Delta Q_2)^2}{4}\right)^2}{N}\\
&+\frac{2\left((\Delta p)^2+\frac{(\Delta P_1)^2}{4}+(\Delta
P_2)^2\right)^2}{N}.
\end{aligned}
\end{equation}
We  note here that the distance measures $d_1$ and $d_2$ for
different measurement schemes are independent of the actual
values of the mean $q_0$ and $p_0$ of the Gaussian states
and depend only on the actual variances $(\Delta q)^2$ and
$(\Delta p)^2$  of the quadratures. Now we turn to study
the dependence of distance measure $d_2$ on the initial
width of the meter $\Delta Q_1$, the squeezing parameter
$r$, and the average number of photons $\langle n\rangle$ 
as we did for $d_1$.

\begin{figure}[h!] \centering
\scalebox{1.0}{\includegraphics{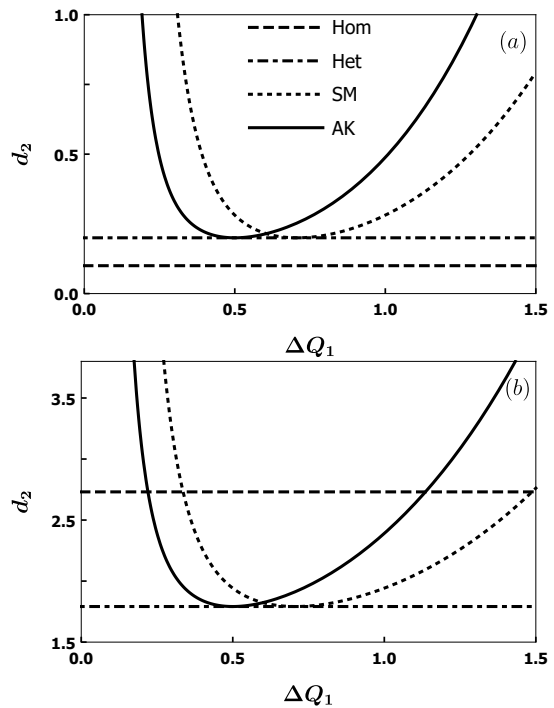}}
\caption{Both the plots show the distance measure $d_2$ as a
function of the initial width of the meter $\Delta Q_1$ for
an ensemble of size $N=20$. $(a)$ The ensemble consists of
identically prepared coherent states.  $(b)$ The ensemble
consists of identically prepared squeezed coherent states
with squeezing parameter $r=1$. } \label{varianceofv}
\end{figure}

We show the plot of the distance measure $d_2$ as a function
of the initial width of the meter $\Delta Q_1$ for a
coherent state ensemble in Fig.~\ref{varianceofv}$(a)$.  The
results show that the homodyne measurement outperforms the
heterodyne measurement in estimating the variance for a
coherent state ensemble. We also notice that the optimal
performance of the Arthurs-Kelly scheme and the sequential
measurement equal the heterodyne measurement. We note that
the optimal performance of the Arthurs-Kelly scheme occurs
at $\Delta Q_1=\Delta P_2=1/2$, while the optimal
performance of the sequential measurement occurs at $\Delta
Q_1=1/\sqrt{2}$.

Similarly, Fig.~\ref{varianceofv}$(b)$ shows the plot of the
distance measure $d_2$ as a function of the initial width of
the meter $\Delta Q_1$ for a squeezed coherent state
ensemble with squeezing parameter $r=1$.  The results show
that the heterodyne measurement outperforms the homodyne
measurement in estimating the variance of a squeezed state
ensemble with squeezing parameter $r=1$. Here too, the
optimal performance of the Arthurs-Kelly scheme and the
sequential measurement equal the heterodyne measurement.

\begin{figure}[h!] \centering
\scalebox{1.0}{\includegraphics{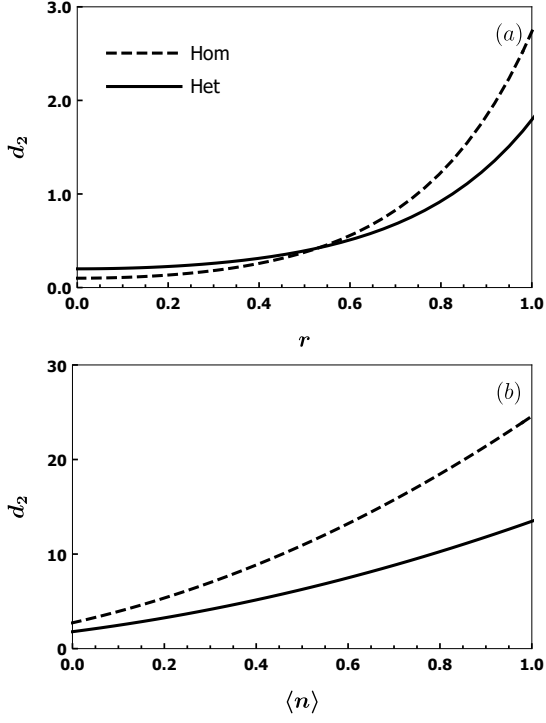}} \caption{ (a)
The distance measure $d_2$
as a function of the squeezing parameter $r$.  Here average
number of photon has been set as	$\langle n \rangle=0$. (b)  The
distance measure $d_2$ as a function of the average photon number
$\langle n \rangle$.  Here squeezing parameter  has been
taken as $r=1$. An ensemble
of size $N=20$ has been 
considered for both the plots. } \label{vvcomp} \end{figure}
We plot the distance measure $d_2$ as a function of the squeezing
parameter $r$ in Fig.~\ref{vvcomp}$(a)$ for $\langle n \rangle=0$. 
The results show
that the homodyne measurement outperforms the heterodyne
measurement up to a certain value of the squeezing parameter
$r_c = 0.53$.  This result is in contrast with the distance measure 
$d_1$ result, where the heterodyne measurement outperforms
the homodyne measurement for all non-zero squeezing
parameter. The critical value of the squeezing parameter
$r_c$ at a given $\langle n \rangle$, 
where the relative performance of the homodyne measurement is equal
to the heterodyne measurement can be written as
\begin{equation}\label{critical}
e^{2r_c} = \frac{1+\sqrt{3+2
n_1^2}+\sqrt{2(2-n_1^2+\sqrt{3+2n_1^2})}}{2 n_1},
\end{equation} 
where $n_1 = 2 \langle n \rangle +1$.  
Further, the plot of the distance measure $d_2$ as a
function of the average number of photons $\langle n\rangle$
is shown in Fig.~\ref{vvcomp}$(b)$. The results reveal that
the variance estimation for a thermal state ensemble is less
precise as compared to a pure state ensemble.

\subsection{Average estimation efficiency}
It is important to know how different schemes perform on the
average. To achieve this we
 compare the relative performances of the
measurement schemes under
consideration on a large number of randomly generated
squeezed coherent thermal states with squeezing parameter
$r$ varying uniformly between $-1$ to $+1$. Such an ensemble
can be produced by a parametric down converter operating at
a fixed temperature, which generates states with squeezing
parameter $r$  uniformly distributed between $-1$ to $+1$.
We note that the squeezing parameter range $-1$ to $+1$ can
be easily achieved in experiments.  For evaluating the
average distance measures $\overline{d_1}$ and
$\overline{d_2}$, we consider the state of the system to be
parameterized by the squeezing parameter $r$ and the average
number of photons $\langle n \rangle$ as given in
Eq.~\eqref{scts}.
\subsubsection{Calculation of  mean distance measure
$\overline{d_1}$}
\begin{figure}[h!]
\centering
\scalebox{1.0}{\includegraphics{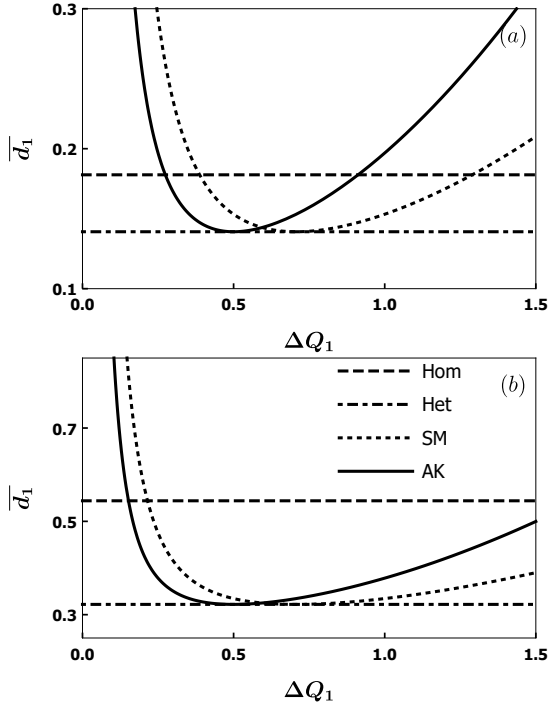}}
\caption{Both the plots show the mean distance measure
$\overline{d_1}$ as a function of the initial width of the
meter $\Delta Q_1$ for an ensemble of size $N=20$. $(a)$ The
averaging is done over identically prepared pure squeezed
coherent state ($\langle n\rangle =0$), whose squeezing
parameter $r$ is uniformly distributed between $-1$ to $+1$.
$(b)$ The averaging is done over identically prepared
squeezed coherent thermal state with $\langle n\rangle =1$,
whose squeezing parameter $r$ is uniformly distributed
between $-1$ to $+1$.}
\label{varaverage}
\end{figure}

The mean distance measure  $\overline{d_1}$ for the
homodyne measurement is calculated as
\begin{equation}
\begin{aligned}
\overline{d_1}^{\mathrm{Hom}}=\frac{1}{2}\int_{-1}^{+1}
d_1^\mathrm{Hom}(r, \langle n\rangle) dr,
=\frac{n_1 \sinh (2)}{N},
\end{aligned}
\end{equation}
where  $n_1=2 \langle n\rangle+1$.
Similarly, the final expressions of the average distance
measure $\overline{d_1}$ for the
other measurement schemes are
\begin{equation}
\begin{aligned}
\overline{d_1}^{\text{Het}}&= \frac{2+n_1 \sinh (2)}{2N},\\
\overline{d}_1^{\text{SM}}&= \frac{2\left( (\Delta
Q_1)^2+(\Delta P_1)^2\right)+n_1 \sinh (2)}{2N},\\
\overline{d_1}^{\text{AK}}&=\frac{1}{4N}\bigg[ (\Delta
Q_2)^2+(\Delta P_1)^2+4\left( (\Delta Q_1)^2+(\Delta
P_2)^2\right)\\
&\quad \quad \quad \quad\quad \quad \quad \quad \quad \quad
\quad\quad\quad \quad+2n_1 \sinh (2)\bigg].
\end{aligned}
\end{equation}
The results for the mean distance measure $\overline{d_1}$ for
various measurement schemes are shown in
Fig.~\ref{varaverage}. 

We see from
Fig.~\ref{varaverage}$(a)$ that the heterodyne measurement
outperforms the homodyne measurement, and the optimal
performance of the Arthurs-Kelly scheme and the sequential
measurement equal the heterodyne measurement.
Figure~\ref{varaverage}$(b)$ shows that the performance
trend for the thermal state ensembles is similar to the pure
state ensembles except for the distance measure of the mean
$\overline{d_1}$ 
is reduced for the thermal state ensemble as compared to the
pure state ensemble.

\subsubsection{Calculation of  mean distance measure $\overline{d_2}$}
We now calculate the expressions of the mean distance measure 
$\overline{d_2}$ averaged over different squeezed coherent
thermal state ensembles with squeezing parameter $r$
uniformly distributed between $-1$ to $+1$ for different
measurement schemes. The mean distance measure $\overline{d_2}$ for
the homodyne measurement and the heterodyne measurement
evaluate to
\begin{equation}
	\begin{aligned}
		& \overline{d_2}^{\mathrm{Hom}}=\frac{n_1^2 \sinh (4)}{2N},\\
		&\overline{d_2}^{\mathrm{Het}}= \frac{4+4n_1 \sinh (2)+n_1^2
			\sinh (4)}{4N}.
	\end{aligned}
\end{equation}
Similarly, the expressions of the mean distance measure $\overline{d_2}$ for the
sequential measurement and the Arthurs-Kelly scheme evaluate to 
\begin{equation}
	\begin{aligned}
		&\overline{d_2}^{\text{SM}}= \frac{1}{4N}\bigg[ 4\left( (\Delta Q_1)^2+(\Delta P_1)^2\right) (2+n_1 \sinh (2))\\
		&\quad\quad\quad \quad\quad\quad \quad\quad\quad \quad\quad\quad \quad\quad\quad \quad+n_1^2 \sinh (4)\bigg],\\
		&\overline{d_2}^{\text{AK}}= \frac{1}{8N}\bigg[\left(4 (\Delta Q_1)^2+(\Delta Q_2)^2+2n_1 \sinh (2)\right)\\
		&\times \left( 4(\Delta Q_1)^2+(\Delta Q_2)^2\right)+\left( (\Delta P_1)^2+4(\Delta P_2)^2\right)  \\
		&\times \left( (\Delta P_1)^2+4(\Delta P_2)^2+2n_1 \sinh (2)\right)+2n_1^2 \sinh (4) \bigg].
	\end{aligned}
\end{equation}

\begin{figure}[h!]
	\centering
	\scalebox{1.0}{\includegraphics{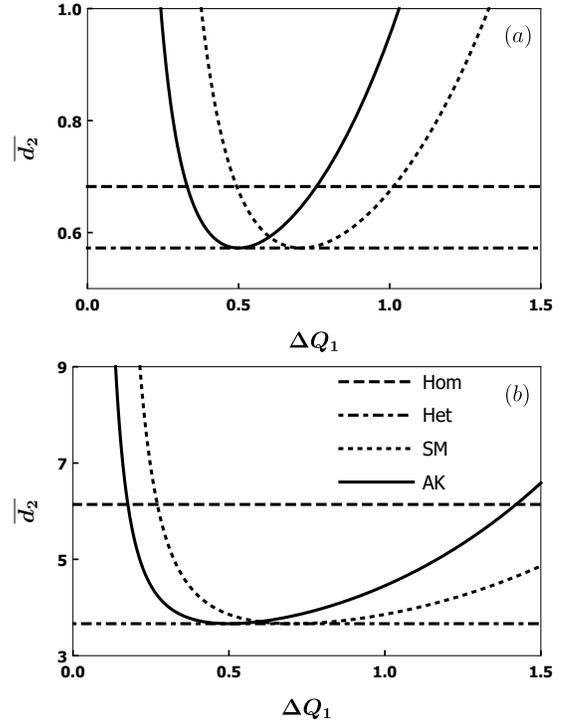}}
	\caption{ Both the plots show the mean distance measure 
		$\overline{d_2}$ as a function of the initial width of the
		meter $\Delta Q_1$ for an	ensemble of size $N=20$.
		$(a)$ The averaging is done over identically prepared
	pure squeezed coherent state ($\langle n\rangle =0$), whose
		squeezing parameter $r$ is uniformly distributed between
		$-1$ to $+1$. $(b)$ The averaging is done over identically
		prepared squeezed coherent thermal state with $\langle
		n\rangle =1$, whose squeezing parameter $r$ is uniformly
		distributed between $-1$ to $+1$. } \label{varvaraverage}
\end{figure}
The results for the mean distance measure $\overline{d_2}$ for
various measurement schemes are shown in
Fig.~\ref{varvaraverage}. 
As can be seen from
Fig.~\ref{varvaraverage}$(a)$, the heterodyne measurement
outperforms the homodyne measurement, and the optimal
performance of the Arthurs-Kelly scheme and the sequential
measurement scheme equal the heterodyne measurement.  For
the thermal state ensembles, the performance trend remains
the same; however,  the  distance measure of 
 the variance $\overline{d_2}$ is reduced for the thermal
state ensembles as compared to the pure state ensembles as
can be seen from Fig.~\ref{varvaraverage}$(b)$.

We summarize the relative performances of the homodyne
measurement and the heterodyne measurement in Table~\ref{table1}.
We further note that the optimal 
performance of the sequential measurement and the 
Arthurs-Kelly scheme is equal to the heterodyne 
measurement for both the mean and the variance estimation.
\begin{table}[h!]
\centering
\caption{\label{table1}
Homodyne measurement versus heterodyne measurement. $d_1$
and $d_2$ represent the accuracy of the mean and the
variance estimation.}
\renewcommand{\arraystretch}{1.5}
\begin{tabular}{ |c |c |c|}
\hline \hline
Ensemble & Distance measure & Rel. performance\\
\hline \hline
Coherent state	($r=0$) & $d_1^{\text{Hom}}$ $=$
$d_1^{\text{Het}}$ & Hom $=$ Het\\ \hline
Squeezed state	($r>0$) & $d_1^{\text{Hom}} >
d_1^{\text{Het}}$ & Hom $<$ Het \\ \hline
$r<r_c$ (Eq.~\ref{critical}) & $d_2^{\text{Hom}}$ $<$
$d_2^{\text{Het}}$ & Hom $>$ Het\\ \hline
$r>r_c$ & $d_2^{\text{Hom}} $ $>$ $ d_2^{\text{Het}}$ & Hom
$<$ Het\\ \hline \hline
$-1 \leq r \leq +1 $ & $\overline{d_1}^{\text{Hom}} >
\overline{d_1}^{\text{Het}}$ & Hom $<$ Het \\  \hline
$-1 \leq r \leq +1 $	&$\overline{d_2}^{\text{Hom}} >
\overline{d_2}^{\text{Het}}$ & Hom $<$ Het \\ \hline
\hline 
\end{tabular}
\end{table}
\section{Modified Hamiltonian in the Arthurs-Kelly scheme}
\label{modifiedak}
Arthurs-Kelly scheme has two measuring probes. What if these
probes can influence each other and are correlated
~\cite{busch-1985}?
To this end, we consider  a modified form of the interaction
Hamiltonian~\cite{busch-1985,bullock-prl-2013}
in the Arthurs-Kelly scheme
\begin{equation}\label{mak}
H =\delta(t-t_1) \left(\hat{q} \hat{P}_1-\hat{p}
\hat{ Q}_2+\frac{\kappa}{2}\hat{P}_1 \hat{Q}_2\right),
\end{equation}
where $\kappa$ determine the coupling strength between
the two probes. This Hamiltonian entangles the system with
both the meters and also the two meters among themselves. 
The corresponding symplectic transformation acting on the 
quadrature operators is
\begin{equation}  
S=\begin{tikzpicture}[baseline=(m-3-1.base),
every left delimiter/.style={xshift=.75em},
every right delimiter/.style={xshift=-.75em}, ]
\matrix [matrix of math nodes,left delimiter=(,right
delimiter=),row sep=0.01cm,column sep=0.01cm] (m) {
1 & 0 & 0 & 0 & -1 & 0 \\
0 & 1 & 0 & -1 & 0 & 0 \\
1 & 0 & 1 & 0 &  \frac{\kappa-1}{2} &0 \\
0 & 0 & 0 & 1 & 0 & 0 \\
0 & 0 & 0 &0  & 1 & 0 \\
0 & 1 & 0 & \frac{-\kappa-1}{2} & 0 & 1 \\};
\draw[solid] ($0.5*(m-1-2.north east)+0.5*(m-1-3.north west)$) --
($0.5*(m-6-3.south east)+0.5*(m-6-2.south west)$);
\draw[solid] ($0.5*(m-1-5.north east)+0.5*(m-1-4.north west)$) --
($0.5*(m-6-4.south east)+0.5*(m-6-5.south west)$);
\draw[solid] ($0.5*(m-2-1.south west)+0.5*(m-3-1.north west)$) --
($0.5*(m-2-6.south east)+0.5*(m-3-6.north east)$);
\draw[solid] ($0.5*(m-4-1.south west)+0.5*(m-5-1.north west)$) --
($0.5*(m-4-6.south east)+0.5*(m-5-6.north east)$);
\node[above=1pt of m-1-1] (top-1) {$q$};
\node[above=1pt of m-1-2] (top-2) {$p$};
\node[above=1pt of m-1-3] (top-3) {$Q_1$};
\node[above=1pt of m-1-4] (top-4) {$P_1$};
\node[above=1pt of m-1-5] (top-5) {$Q_2$};
\node[above=1pt of m-1-6] (top-6) {$P_2$};
\node[right=2pt of m-1-6] (left-1) {$q$};
\node[right=2pt of m-2-6] (left-1) {$p$};
\node[right=2pt of m-3-6] (left-1) {$Q_1$};
\node[right=2pt of m-4-6] (left-1) {$P_1$};
\node[right=2pt of m-5-6] (left-1) {$Q_2$};
\node[right=2pt of m-6-6] (left-1) {$P_2$};
\end{tikzpicture},
\end{equation}

The covariance matrix and the displacement vector
corresponding to system-meters state after time $t_1$ 
can be evaluated using Eq.~(\ref{transformation}). The 
covariance matrix of the reduced state of the two meters 
is given by $V_{M_1 M_2}^{\mathrm{RED}}=   $
\begin{equation}\label{covaraincereduced}
 \begin{psmallmatrix}
V_{M_1}(Q_1) & 0 & \frac{(\kappa-1)}{2}(\Delta Q_2)^2 & 0 \\
0 & (\Delta P_1)^2 &0 &  -\frac{(\kappa+1)}{2}(\Delta P_1)^2 \\
\frac{(\kappa-1)}{2}(\Delta Q_2)^2 & 0 & (\Delta Q_2)^2 & 0\\
0&-\frac{(\kappa+1)}{2}(\Delta P_1)^2 & 0& V_{M_2}(P_2) \\
\end{psmallmatrix},
\end{equation}

where
\begin{equation}
\begin{aligned}
V_{M_1}(Q_1)=&(\Delta q)^2+(\Delta Q_1)^2+
\frac{(\kappa-1)^2}{4}(\Delta Q_2)^2,\\
V_{M_2}(P_2)=&(\Delta p)^2+\frac{(\kappa+1)^2}{4}
(\Delta P_1)^2+(\Delta P_2)^2.
\end{aligned}
\end{equation}
We find, using Simon's entanglement criteria~\cite{simon-prl-2000},
that the reduced state of the two meters is  entangled for $|\kappa|\geq 1$.
The variance of the probability distribution for the measurement
of the $\hat{q}$-quadrature on the meter 1 and the $\hat{p}$-quadrature 
on the meter 2 can be written as the variance corresponding
to $\hat{Q}_1$ and $\hat{P}_2$ in the covariance matrix 
for the reduced state of the meters~(\ref{covaraincereduced}):
\begin{equation}
\begin{aligned}
V^{\mathrm{COR}}(\hat{q}) = & (\Delta q)^2+(\Delta Q_1)^2+
\frac{(\kappa-1)^2}{4}(\Delta Q_2)^2,\\
V^{\mathrm{COR}}(\hat{p}) = & (\Delta p)^2
+\frac{(\kappa+1)^2}{4}(\Delta P_1)^2+(\Delta P_2)^2.
\end{aligned}
\end{equation}
Thus, the distance measure $d_1$ for the 
modified Arthurs-Kelly scheme reads
\begin{equation}
d_1^{\mathrm{COR}}=\frac{V^{\mathrm{COR}}(\hat{q})}{N}
+\frac{V^{\mathrm{COR}}(\hat{p})}{N}.
\end{equation}
We optimize the distance measure $d_1^{\mathrm{COR}}$ with
respect to the parameters $\Delta Q_1$ and $\Delta P_2$.
The optimal value of the distance measure
$d_1^{\mathrm{COR}}$  evaluates to
\begin{equation}
d{_{1\,{\mathrm{OPT}}}^{\mathrm{COR}}} =
\begin{cases}
\frac{1+(\Delta q)^2+(\Delta p)^2}{N} & |\kappa| \leq1, \\
\frac{|\kappa|+(\Delta q)^2+(\Delta p)^2}{N} &  |\kappa| >1. \\
\end{cases}
\end{equation}
We note that the optimal  distance measure 
$d{_{1\,{\mathrm{OPT}}}^{\mathrm{COR}}}$ for 
$|\kappa| \leq1$ equals the distance measure for the
heterodyne measurement $d_1^{\mathrm{Het}}$~(\ref{vhetero}).

\begin{figure}[h!]
\centering
\scalebox{1.0}{\includegraphics{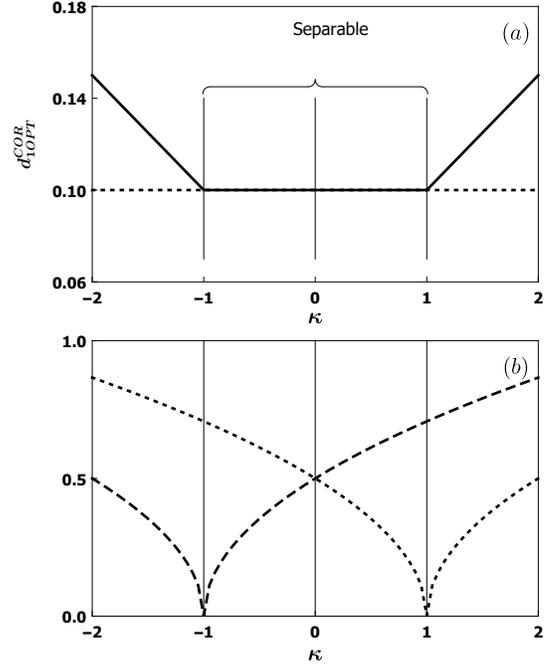}}
\caption{(a) Optimal distance measure
$d{_{1\,{\mathrm{OPT}}}^{\mathrm{COR}}}$ for the modified
Arthurs-Kelly scheme, represented by the solid curve, as a
function of the coupling strength $\kappa$.  The dashed
curve represents the distance measure  for the heterodyne measurement
$d_1^{\mathrm{Het}}$.  (b) The plot of $\Delta Q_1$ (dashed)
and $\Delta P_2$ (dotted) for the optimal performance of the
modified Arthurs-Kelly scheme as a function of the coupling
strength $\kappa$.  We have considered an ensemble of
coherent states for both the plots. } \label{varcor}
\end{figure}

The plot of the distance measure $d_1$ as a function of the coupling
strength $\kappa$ for a coherent state ensemble is shown in
Fig.~\ref{varcor}(a). 
The results show that the estimation
of the coherent state ensemble using the modified
Arthurs-Kelly scheme is best in the range $|\kappa| \leq1$,
which corresponds to uncorrelated probes.  The corresponding
value of $\Delta Q_1$ and $\Delta P_2$, which optimizes the
distance measure $d_1^{\mathrm{COR}}$ turns out to be
\begin{equation}
\Delta Q_1 =
\begin{cases}
\frac{\sqrt{1+\kappa}}{2} & \kappa > 1, \\
\frac{\sqrt{1+\kappa}}{2} &  |\kappa| <1, \\
\frac{\sqrt{-1-\kappa}}{2} &  \kappa < -1, \\
\end{cases}
\quad \Delta P_2 =
\begin{cases}
\frac{\sqrt{-1+\kappa}}{2} & \kappa > 1, \\
\frac{\sqrt{1-\kappa}}{2} &  |\kappa| <1, \\
\frac{\sqrt{1-\kappa}}{2} &  \kappa < -1. \\
\end{cases}
\end{equation}
We have also plotted  $\Delta Q_1$ and $\Delta P_2$ as a
function of the coupling strength $\kappa$ corresponding to
the optimal performance of the modified Arthurs-Kelly scheme
in Fig.~\ref{varcor}(b).  For the coupling strength
$\kappa=0$, the modified Arthurs-Kelly  scheme reduces to
the original Arthurs-Kelly scheme. This can also be verified
from Fig.~\ref{varcor}(b), where at $\kappa=0$, $\Delta
Q_1 = \Delta P_1 = 1/2$, which is the same as
Eq.~(\ref{optimal}).

  Furthermore, the analysis for the distance measure $d_2$ also shows that the estimation
of the coherent state ensemble using the modified
Arthurs-Kelly scheme is best in the range $|\kappa| \leq1$.

\section{Discussion and Conclusion}
\label{conclusion}
In this paper, we have explored the
estimation  of the mean and  the variance of an ensemble
of a fixed number of identically prepared
Gaussian states by employing four
different  measurement schemes with a view to compare their
efficiencies. Since we were dealing with
Gaussian states and quadratic Hamiltonians, the covariance
matrix, phase space formulation and symplectic group
techniques  provided an elegant
and intuitive way to handle the analysis. Detailed analysis of the
distance measures revealed that the optimal performance of
the Arthurs-Kelly scheme requires non-classical resources in
the sense that the meters should be initially prepared in a
squeezed state; however, the optimal performance of the
sequential measurement only requires classical resources,
\ie, the meter should be initially prepared in a coherent
state. 
Further, we showed that the optimal performance of
the Arthurs-Kelly scheme and the sequential measurement
equal heterodyne measurement for both the mean and the
variance estimation. 

 For mean estimation, the analysis
revealed that the performance of the homodyne measurement
and the heterodyne measurement is the same for a coherent
state ensemble, whereas, for a squeezed state ensemble, the
heterodyne measurement performs better than the homodyne
measurement.  
For variance estimation, the homodyne
measurement outperforms the heterodyne measurement for a
squeezed coherent thermal state ensemble  up to a certain
squeezing parameter range.  
The results show that the heterodyne
measurement always perform better than the homodyne
measurement for both the mean and the variance estimation on
the average.
We considered the possibility of correlated probes for 
Arthurs-Kelly scheme and showed that optimal performance of the scheme
can  only be obtained  when the meters are uncorrelated.

We expect that these results will find applications in
various quantum information and quantum communication
protocols. One natural extension that we are pursuing is to
extend the analysis for Gaussian states squeezed in
arbitrary directions. It would be interesting to generalize
the theory for non-Gaussian states, where we
are required to estimate higher order moments.
\section*{Acknowledgement}
C. K. thanks Shikhar Arora for his crucial remarks on the
final version of the manuscript.  A and C.K. acknowledge the
financial support from {\sf
DST/ICPS/QuST/Theme-1/2019/General} Project number {\sf
Q-68}.

\appendix
\section{Calculation of the symplectic transformation matrix
for a given Hamiltonian}
\label{symplecticcalculation}
We provide two different methods to evaluate the symplectic
transformation matrix for a given Hamiltonian.
\subsection*{Method I: Hilbert space and
Baker-Campbell-Hausdorff formula}
Consider the Hamiltonian  $ \hat{H}(t) =
\delta(t-t_1)\hat{q}\hat{P_1}$ (Eq.~\ref{intseq}). The
corresponding infinite dimensional unitary operator for
$t>t_1$ is given by
\begin{equation}
\mathcal{U}(\hat{H}(t)) =\mathrm{e}^{-i\int \hat{H}(t) \,dt}
=\mathrm{e}^ { -i\hat{q}\hat{P_1} }.
\end{equation}
In Heisenberg picture, the evolution of any operator
$\hat{A}$ can be written as
\begin{equation}
\hat{A}\xrightarrow{\mathcal{U}(\hat{H}(t))}\mathcal{U}(\hat{H}(t))^{\dagger}\,
\hat{A}\,\mathcal{U}(\hat{H}(t)).
\end{equation}
Thus, the transformation of various quadrature
operators using the Baker-Campbell-Hausdorff (BCH) formula
can be evaluated as following:
\begin{equation}
\begin{aligned}
\mathrm{e}^ { i\hat{q}\hat{P}_1 }\, \hat{q}\,\mathrm{e}^ { -i\hat{q}\hat{P}_1 }&=\hat{q}\\
\mathrm{e}^ { i\hat{q}\hat{P}_1 }\, \hat{p}\,\mathrm{e}^ { -i\hat{q}\hat{P}_1 }&=\hat{p}-\hat{P}_1\\
\mathrm{e}^ { i\hat{q}\hat{P}_1 }\, \hat{Q_1}\,\mathrm{e}^ { -i\hat{q}\hat{P}_1 }&=\hat{q}+\hat{Q}_1\\
\mathrm{e}^ { i\hat{q}\hat{P}_1 }\, \hat{P_1}\,\mathrm{e}^ { -i\hat{q}\hat{P}_1 }&=\hat{P_1}\\
\end{aligned}
\end{equation}
Thus, the quadrature operators transform as
\begin{equation}\label{hilbert}
\begin{pmatrix}
\hat{q}\\
\hat{p}\\
\hat{Q}\\
\hat{P_1}
\end{pmatrix}
\xrightarrow{\mathcal{U}(\hat{H}(t))}
\underbrace{ \left(
\begin{array}{cccc}
1 & 0 & 0 & 0 \\
0 & 1 & 0 & -1 \\
1 & 0 & 1 & 0 \\
0 & 0 & 0 & 1 \\
\end{array}
\right)}_{S}
\begin{pmatrix}
\hat{q}\\
\hat{p}\\
\hat{Q}\\
\hat{P_1}
\end{pmatrix},
\end{equation}
where $S$ is the symplectic transformation corresponding to
the Hamiltonian $ \hat{H}(t) =
\delta(t-t_1)\hat{q}\hat{P_1}$.
However, this method gets a 
little complicated for the modified Arthurs-Kelly scheme,
where the Hamiltonian is $H =\delta(t-t_1) \left(\hat{q} \hat{P}_1-\hat{p}
\hat{ Q}_2+\frac{\kappa}{2}\hat{P}_1 \hat{Q}_2\right)$. 
Alternatively, we can take another approach where such complicated
calculations can be performed  easily.
\subsection*{Method II: Exponentiation of the generators of
$Sp(2n,\mathcal{R})$}
Let $J$ be the generator of the symplectic group
$Sp(2n,\mathcal{R})$, \ie, $J$ is an element of the Lie
algebra of  $Sp(2n,\mathcal{R})$ group.
The corresponding symplectic group element $S$ can be
obtained by exponentiating $J$ as follows:
\begin{equation}
S= \exp(J).
\end{equation}
We can associate a quadratic function
of quadrature operators with every $J$,  which is Hermitian,
as follows:
\begin{equation}
H(J) = \frac{1}{2}\hat{\xi}^T (\Omega J )\hat{\xi},
\end{equation}
where $\hat{\xi}$ is the column of quadrature operators and
$\Omega$ is the symplectic form.
Since the generators of the symplectic group and quadratic
functions of the quadrature operators are in one-to-one
correspondence at Lie algebra level, we can exponentiate
$H(J)$ to obtain infinite-dimensional unitary representation
of $ S= \exp(J)$. Thus, in our case, we can first determine
the generator $J$ from the given Hamiltonian and evaluate
the corresponding symplectic transformation by
exponentiation.  We illustrate this procedure for
of the Hamiltonian $ \hat{H}(t) =
\delta(t-t_1)\hat{q}\hat{P_1}$, whose corresponding infinite
dimensional unitary representation is $\mathrm{e}^ {
-i\hat{q}\hat{P_1} }$. We can write
\begin{equation}
-i\hat{q}\hat{P_1} =  \frac{1}{2}\hat{\xi}^T (\Omega J) \hat{\xi} ,
\end{equation}
where $  \hat{\xi} = (\hat{q}, \,\hat{p},\,
\hat{Q}_{1},\,\hat{P}_{1})^{T}$ and 
\begin{equation}
\Omega J  = -\left(
\begin{array}{cccc}
0 & 0 & 0 & 1 \\
0 & 0 & 0 & 0 \\
0 & 0 & 0 & 0 \\
1 & 0 & 0 & 0 \\
\end{array}
\right).
\end{equation}
Consequently, the generator $J$ becomes
\begin{equation}
J=  \left(
\begin{array}{cccc}
0 & 0 & 0 & 0 \\
0 & 0 & 0 & -1 \\
1 & 0 & 0 & 0 \\
0 & 0 & 0 & 0 \\
\end{array}
\right).
\end{equation}
Thus, the symplectic matrix corresponding to the generator $J$ is   
\begin{equation}
S=\exp(J)=      \left(
\begin{array}{cccc}
1 & 0 & 0 & 0 \\
0 & 1 & 0 & -1 \\
1 & 0 & 1 & 0 \\
0 & 0 & 0 & 1 \\
\end{array}
\right),
\end{equation}
which is the same as the symplectic transformation matrix
obtained using the BCH formula in the previous section.

\begin{widetext}
\section{Final state in Arthurs-Kelly scheme }\label{appendixAK}
The displacement vector and the covariance matrix of the
transformed joint system-meters state is given by
\begin{equation}\label{appendixdis}
\overline{\hat{\xi}^{'}}=   \begin{pmatrix}
\langle \hat{q}\rangle = q_0 \\
\langle \hat{p}\rangle = p_0 \\
\langle \hat{Q_1}\rangle = q_0 \\
\langle \hat{P_1}\rangle = 0  \\
\langle \hat{Q_2}\rangle = 0  \\
\langle \hat{P_2}\rangle = p_0 \\
\end{pmatrix},
\end{equation}
and
\begin{equation} \label{appendixcov} 
V^{'}=\begin{tikzpicture}[baseline=(m-3-1.base),
every left delimiter/.style={xshift=.75em},
every right delimiter/.style={xshift=-.75em},
]
\matrix [matrix of math nodes,left delimiter=(,right
delimiter=),row sep=0.01cm,column sep=0.01cm] (m) {
(\Delta q)^2+ (\Delta Q_2)^2 & 0 &(\Delta q)^2
+\frac{(\Delta Q_2)^2}{2} & 0 & -(\Delta Q_2)^2 & 0 \\
0 &(\Delta p)^2+ (\Delta P_1)^2& 0 & -(\Delta P_1)^2 & 0&
(\Delta p)^2+\frac{(\Delta P_1)^2}{2}  \\
(\Delta q)^2 +\frac{(\Delta Q_2)^2}{2}& 0 &(\Delta
q)^2+(\Delta Q_1)^2+ \frac{(\Delta Q_2)^2}{4} & 0 &
-\frac{(\Delta Q_2)^2}{2} & 0 \\
0 & -(\Delta P_1)^2& 0 & (\Delta P_1)^2 &0 &  -\frac{(\Delta
P_1)^2}{2} \\
-(\Delta Q_2)^2 & 0 & -\frac{(\Delta Q_2)^2}{2} & 0 &
(\Delta Q_2)^2 & 0\\
0 &(\Delta p)^2+ \frac{(\Delta P_1)^2}{2} & 0 &
-\frac{(\Delta P_1)^2}{2} & 0& (\Delta p)^2+\frac{(\Delta
P_1)^2}{4}+(\Delta P_2)^2 \\
};

			
			
\node[above=1pt of m-1-1] (top-1) {$q$};
\node[above=1pt of m-1-2] (top-2) {$p$};
\node[above=1pt of m-1-3] (top-3) {$Q_1$};
\node[above=1pt of m-1-4] (top-4) {$P_1$};
\node[above=1pt of m-1-5] (top-5) {$Q_2$};
\node[above=1pt of m-1-6] (top-6) {$P_2$};

\coordinate[above=3pt of top-1] (aux);
\begin{scope}[nodes={text depth=0.25ex}]
\draw[thick,decorate,decoration=brace] (top-1.west|-aux) -- (top-2.east|-aux)
node[midway,above=0.1ex] {System};
\draw[thick,decorate,decoration=brace] (top-3.west|-aux) -- (top-4.east|-aux)
node[midway,above=0.1ex] {Meter 1};
\draw[thick,decorate,decoration=brace] (top-5.west|-aux) -- (top-6.east|-aux)
node[midway,above=0.1ex] {Meter 2};
\end{scope}


\end{tikzpicture}.
\end{equation}
\end{widetext}
%

\end{document}